\documentclass[trackchanges,twocolumn]{aastex701}

\usepackage{graphicx}
\usepackage{subfig}
\usepackage{scrextend}
\usepackage{amsmath}
\usepackage{gensymb}

\begin{document}

\title{AGN versus Star-formation: A MUSE Analysis of NGC 1365}

\author[orcid=0009-0008-2653-8446,sname='Mullaney']{Kyla Mullaney}
\affiliation{X-ray Astrophysics Laboratory, NASA Goddard Space Flight Center, Code 662, Greenbelt, MD 20771, USA}
\affiliation{Southeastern Universities Research Association (SURA)}
\affiliation{Center for Research and Exploration in Space Science and Technology, NASA/GSFC, Greenbelt, MD 20771}
\email[show]{kylamullaney@gmail.com}  

\author[orcid=0000-0002-8571-9801,gname=Kelly, sname='Whalen']{Kelly E. Whalen} 
\affiliation{X-ray Astrophysics Laboratory, NASA Goddard Space Flight Center, Code 662, Greenbelt, MD 20771, USA}
\email{kelly.e.whalen@nasa.gov}

\author[orcid=0000-0003-1051-6564,gname=Cann,sname=Jenna]{Jenna Cann}
\affiliation{X-ray Astrophysics Laboratory, NASA Goddard Space Flight Center, Code 662, Greenbelt, MD 20771, USA}
\affiliation{Center for Space Science and Technology, University of Maryland, Baltimore County, 1000 Hilltop Circle, Baltimore, MD 21250, USA}
\affiliation{Center for Research and Exploration in Space Science and Technology, NASA/GSFC, Greenbelt, MD 20771}
\email{jenna.cann@nasa.gov}

\author[orcid=0009-0008-4232-486X,gname=Kimberly,sname=Weaver]{Kimberly Weaver}
\affiliation{X-ray Astrophysics Laboratory, NASA Goddard Space Flight Center, Code 662, Greenbelt, MD 20771, USA}
\email{kimberly.a.weaver@nasa.gov}

\author[0000-0002-0913-3729]{Jeffrey McKaig}
\affiliation{X-ray Astrophysics Laboratory, NASA Goddard Space Flight Center, Code 662, Greenbelt, MD 20771, USA}
\affiliation{Oak Ridge Associated Universities, NASA NPP Program, Oak Ridge, TN 37831, USA}
\email{jeffrey.d.mckaig@nasa.gov}

\author[orcid=0000-0003-1051-6564,gname=Doan,sname=Sara]{Sara Doan}
\affiliation{Department of Physics and Astronomy, George Mason University, Fairfax, VA, 22030}
\email{sdoan2@gmail.com}

\begin{abstract}

Active galactic nuclei (AGN) and star formation feedback may heat and remove gas from galaxies in a process that quenches ongoing star formation and shapes the evolution of galaxies. Potential impacts from these processes can be seen in the complex and interconnected signatures of AGN and star formation activity throughout a galaxy. Here, we analyze archival integral field unit (IFU) data for the nearby Seyfert galaxy, NGC 1365, as observed with the Multi Unit Spectroscopic Explorer (MUSE) instrument on the Very Large Telescope (VLT). Our analysis probes the ionization and kinematic properties of NGC 1365 at high spatial resolution over unprecedentedly large physical scales (approximately 40 kpc), allowing us to trace the effects of feedback throughout nearly an entire galaxy. We use these optical IFU data in conjunction with observations from the James Webb Space Telescope (JWST) and Chandra X-ray Observatory to analyze and compare maps of emission line flux, ionization state, star formation, and gas kinematics. In doing so, we identify a region of BPT-identified unexpectedly high ionization relative to surrounding areas in the star forming arms, and work to identify its source, finding that shock heating may play a significant role. Results from this analysis allow us to place constraints on the relative impact of AGN and star formation processes on the star forming gas in NGC 1365, as well as begin to inform our understanding on the global impacts of feedback in galaxy populations as a whole.

\end{abstract}

\keywords{\uat{Galaxies}{573} --- \uat{Galaxy evolution}{594} --- \uat{Active galactic nuclei}{16} --- \uat{Star formation}{1569}}

\section{Introduction} \label{sec:introduction}

Understanding the co-evolution of galaxies and their central supermassive black holes (SMBHs) is critical for a deeper understanding of the history and progression of structure formation in the universe. Evidence for an intrinsic link between the evolution of a SMBH and its host galaxy can be seen in the establishment of several scaling relations between the mass and accretion activity of the SMBH and larger galaxy properties \citep{AlexanderHickox2025} - such as the velocity dispersion \citep{Ferrarese2000Typhoon, Gebhardt2000Typhoon}, stellar mass, and luminosity of the bulge \citep{Magorrian1998Typhoon, Marconi2003Typhoon}. However, these relationships can become complicated when considering wider ranges of galaxy and black hole mass, as well as different regions of the host galaxy. In a review assessing the accuracy of scaling relations for several surveys and mass derivation techniques, \citet{ORNOTreview2013} find significant variation in the correlation strength of SMBH-host galaxy scaling relations, depending on galaxy structure and components, as well as galaxy masses. Furthermore, the relationship between SMBH accretion activity and galactic star formation rates, a key part of galaxy evolution, is not entirely clear \citep{AlexanderHickox2012Typhoon}. 

Feedback processes --- spurred by shocks or winds from star formation, actively accreting supermassive black holes (active galactic nuclei; AGN), or supernovae --- inject energy into the surrounding ISM and constitute one avenue by which a central SMBH and its host galaxy interact. This may heat or expel molecular gas, limiting the ability of a galaxy to form stars by preventing the clumping of cold gas needed to fuel star formation (SF) \citep{Noeske2007Lomaeva, Wuyts2011Lomaeva}. However, the relationships between AGN- and SF-driven feedback processes, and their respective effects on a host galaxy, are unclear \citep{ORNOTreview2013}. For example, AGN-induced feedback can decrease star formation rates (SFRs) by driving outflows that heat or remove gas \citep{Croton2006Lomaeva, Heckman2014Lomaeva}, or may increase SFRs by creating shocks in the ISM, increasing the ability of gas to clump and form stars in certain areas \citep{Silk2010Lomaeva}. 

This ambiguity is further complicated by the presence of factors outside of feedback processes (e.g., environmental quenching via ram pressure stripping, morphological factors such as bars or bulges) that may significantly spur or quench star formation as well. Moreover, observational \citep[see review by][]{ObservationReview2024} and theoretical \citep{Weinberger2018AH, Terrazas2020AH, Wellons2023AH} studies of AGN-driven feedback have suggested its potential necessity in driving the bimodal distribution of galaxies, between bluer, typically younger, galaxies undergoing active SF and redder, typically older, ``quenched" elliptical galaxies. However, there exists no clear-cut understanding of the mechanism by which AGN-driven feedback affects its host galaxy on such wide physical and temporal scales \citep{ObservationReview2024}. 

\citet{AlexanderHickox2025} suggests that one of the primary developments in the last decade of research in this field is a shift from the attempt to find direct, population level evidence of AGN feedback, towards a combined strategy of assessing the smaller physical scale mechanisms of feedback via detailed views of resolved data, especially utilizing integral field unit (IFU) instruments, and addressing the population level understanding of the role of feedback through larger survey studies. This work addresses this stated need for high spatial resolution study of feedback mechanisms through a detailed study of NGC 1365.

NGC 1365 is a nearby ($\approx18$ Mpc), massive ($\approx10^{10.75}$~M$_\odot$, \citep{Leroy2019}) dusty barred spiral (SB(s)b) galaxy \citep{deVaucoulers1991MIRACLE} displaying both AGN and starburst activity \citep[classed as a Seyfert 1.8][]{Veron2006MIRACLE}. NGC 1365's proximity and wide observational coverage provides a strong case study for in-depth analysis of the interplay between AGN- and SF-driven feedback processes at both large and small scales. The dynamic environment of the galaxy is complex, hosting a circumnuclear starburst ring obscured by a nuclear dust ring \citep{Alonso2012MIRACLE}, as well as a changing-look AGN that drives a bi-conical outflow \citep{Venturi2018MIRACLE, Risaliti2005Fazeli, Braito2014Fazeli}.

The galaxy has a SFR of 16.9 M$_\odot$ per year \citep{Whitmore_2023}, with the majority of star formation concentrated within the circumnuclear ring \citep{Alonso2012MIRACLE}. Additionally, a comparison of data from \textit{JWST} and \textit{ALMA} uncovers regions of decreased excitation and increased dissociation in molecular gas, interpreted as being affected by stellar feedback from nuclear-region young massive stellar clusters \citep{Liu2022}. 
The AGN has been classified as a changing look AGN (CLAGN), based on variability in the column density of the X-ray absorber using XMM-Newton time-resolved spectra \citep{Risaliti2005Fazeli, Braito2014Fazeli}. Morphologically, NGC 1365 contains a long bar with a length of $\sim$3 arcminutes, causing an inflow of gas \citep{ReganandElmegreen1997Fazeli}, and a dust lane. NGC 1365 is located in the Fornax cluster \citep{JonesandJones1980MIRCALE}, at a redshift of $z = 0.005457$ \citep{Bureau1996MIRACLE}, and we adopt a redshift independent mean distance of 17.83 $\pm$ 0.51 Mpc \citep{Willick2001}.

We present the first very high resolution (post-binning 0.9-1.2 arcsecond per spatial/spectral pixel (spaxel); approximately 105-140 pc/spaxel) analysis of kinematics and ionization in the star-forming arms of NGC 1365. In Section \ref{sec:data}, we identify our archival data sources and describe our IFU spectral fitting methodology. In Section \ref{sec:results}, we present spatially-resolved maps of the ionization source, AGN-SF mixing fraction, SFR surface density, and [OIII]$\lambda$5007 kinematics, and analyze the relationships between kinematic and star formation markers in AGN- and SF- dominated regions of the galaxy. We additionally identify a region of high-ionization in the star forming arms. In Section \ref{sec:discussion}, we discuss interpretations for the relative contributions of AGN- and SF-driven feedback, identify potential sources of the high-ionization region in the star forming arms, and note implications for potential future studies. In Section \ref{sec:conclusion}, we report our conclusions. 

\section{Data and Methods}
\label{sec:data}

\subsection{Data}
\label{subsec:data_only}

We utilize an archival mosaic image (Program ID 094.B-0321; PI Marconi, Schinnerer) from the Multi Unit Spectroscopic Explorer (MUSE) instrument on the Very Large Telescope (VLT) \citep{MUSE}, accessed via the ESO Science Data Archive\footnote{https://archive.eso.org/scienceportal/home}. 
The mosaic has a field of view of $5.677$\arcmin, covering the central nuclear region and central portion of the star-forming arms. The observations were obtained on October 12, 2014 in the wide-field mode without adaptive optics. The spectral range is 475-935 nm with an average spectral resolution of $R\approx3000$. The effective exposure time is 33.4 ks and the average seeing was 0.\arcsec823.

We additionally utilize archival mid-infrared observations from Mid-InfraRed Instrument (MIRI) on \textit{JWST}, sourced from the Mikulski Archive for Space Telescopes (MAST) database\footnote{https://mast.stsci.edu/ \label{fn:mast_footnote}}. 
The data were obtained on August 13, 2022 (PI Lee, Program ID GO \#2107). Data were collected using MIRI imaging mode with the F2100W ($21\mu$m) filter. The observations used the FULL subarray, the FASTR1 readout pattern, and a 4-point dither pattern optimized for extended sources. The full exposure time was 1287 s, with 14 groups per integration and two integrations per exposure at each of the four dither positions. 
For calibration purposes for the \textit{JWST} data, archival Spitzer MIPS (Multiband Imaging Photometer for Spitzer) data was sourced from the Spitzer Heritage Archive\footnote{https://irsa.ipac.caltech.edu/applications/Spitzer/SHA}) (Program ID 3672; PI Mazzarella). The Spitzer data were obtained on December 24, 2004 using the 24$\mu$m band. 

Finally, archival Chandra observations (PI Risaliti, ObsIds: 6868, 6869, 6870, 6871, 6872, 6873) were used to identify sources of high-energy emission for comparison with multi-wavelength datasets. These data were downloaded from the Chandra Data Archive\footnote{https://cda.harvard.edu/chaser/}, and reprocessed and combined into a single events file using the $\verb|chandra_repro|$ and $\verb|reproject_obs|$ commands from CIAO v4.17 (Chandra Interactive Analysis of Observations).

\subsection{Spectral Fitting}
\label{subsec:ifu_fitting}

We perform spectral fitting on the MUSE mosaic using the software Bayesian AGN Decomposition Analysis for SDSS Spectra \citep[BADASS; ][]{BADASS}. BADASS is an open source software for spectral analysis, initially intended for use on SDSS spectra, but later adapted for parallelized analysis of IFU data \citep{Reefe_2023, Doan_2025, Aravindan_2023, Bohn_2024}. It uses the Markov Chain Monte Carlo emcee package \citep{emcee} and autocorrelation analysis to perform simultaneous fitting of spectral components including narrow, broad, and outflow emission line features, as well as power-law continuum, stellar line-of-sight velocity distribution, and Fe~II emission. We perform spectral fitting on the MUSE image using BADASS, with a fixed bin size of 3 square spaxels to match seeing resolution, to obtain maps of emission line flux, SNR (signal-to-noise ratio), and associated kinematic information, including the velocity dispersion, full-width half-maximum (FWHM), and velocity offset, for features of interest, particularly [\ion{O}{3}]$\lambda$5007, [\ion{N}{2}]$\lambda$6585, [\ion{S}{2}]$\lambda$6718,6732, [\ion{O}{1}]$\lambda$6302, H$\alpha$$\lambda$6563, and H$\beta$$\lambda$4862. 
We note that we use only single-component Gaussian fits to calculate narrow line fluxes and bulk velocities. For more in-depth kinematic analyses, we reference the nuclear region analysis in \citet{Venturi2018MIRACLE}. We focus in this work on expanding the view of the ionization state to the star forming arms.

\begin{figure}[h!]
\includegraphics[width=0.50\textwidth]{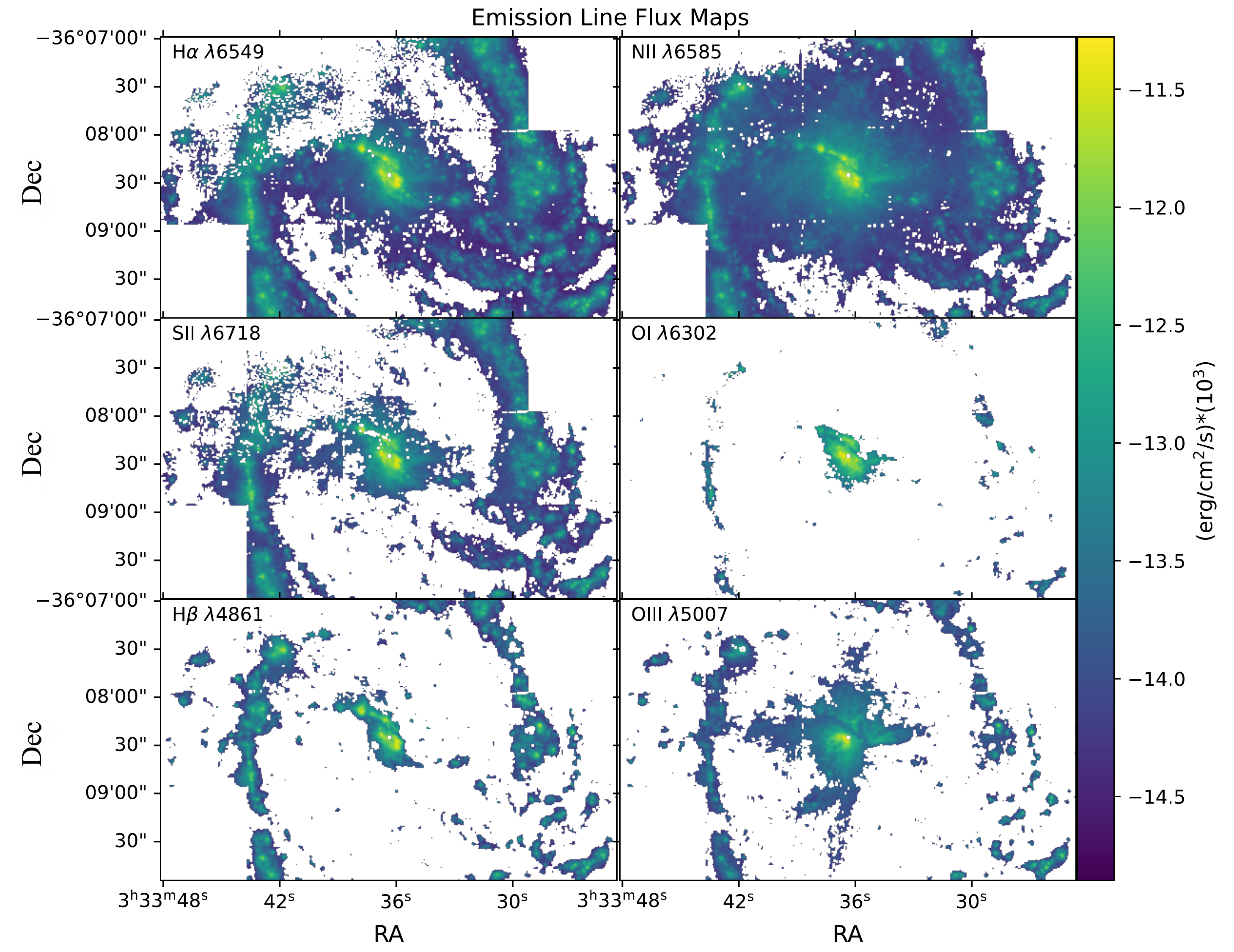}
\caption{Flux maps of the primary 
emission lines used in this work, computed from the maximum likelihood BADASS models. All are masked as specified for each line flux, requiring a 3$\sigma$ detection, an SNR greater than 8, percent residual error $< 20$\%, and a velocity offset $< 400$~km~s$^{-1}$. These masks are used throughout the work, and combined when a figure references more than one emission line. On the left, moving from top to bottom, we show the the fluxes for H$\alpha\lambda6563$, [\ion{S}{2}]$\lambda$6718, H$\beta\lambda4861$, and on the right from top to bottom, we show the flux for [\ion{N}{2}]$\lambda$6585, [\ion{O}{1}]$\lambda$6302, [\ion{O}{3}]$\lambda$5007.
\label{fig:6_panelfluxmaps}}
\end{figure}

In our analysis, we utilize maps of emission line flux and kinematic information (see Figure \ref{fig:6_panelfluxmaps} for example flux maps). We apply emission-line-specific masks to each individual flux and kinematic map to ensure we only include spaxels that have high-quality spectral fits. Due to the inherent challenge of fitting extremely large and highly spatially resolved IFU samples, we expect the presence of occasional spurious fitting errors, and utilize four stringent masks based on the SNR, standard deviation noise, velocity offset, and fit residual to eliminate as many spurious fit errors as possible. Additionally, while this study does not require the robust fitting of broad and outflow components, the potential presence of these features in select spaxels could result in over- or under-estimation of the calculated fluxes. We define our masks to account for this added uncertainty.
In constructing these masks, we only include those spaxels with an SNR $>$ 8 for the given emission line. We further implement a cut using the program-determined standard deviation, removing any spaxels with fluxes that do not meet a  3$\sigma$ detection criteria for the emission line in question. To account for spurious fits of nearby features or noise, we additionally remove any spaxels with a velocity offset for a given emission line of larger than 400 km/s. Finally, we required the integrated residuals over the relevant bin covering the full width of the emission line, accounting for rotation ($\approx15\AA$) 
to be $<20\%$ of the fit data. 
As a final check, we visually inspected significant outliers in FWHM and emission line diagnostic diagrams and manually masked out poor fits. Example flux maps are shown in figure \ref{fig:6_panelfluxmaps}. We use these to construct Baldwin, Philips, and Terlevich (BPT) diagrams \citep{BPT} and kinematic maps later in this analysis, and mask these by combining the masks of each individual emission line used in a given metric.

\section{Results} \label{sec:results}

\subsection{Star-formation Rates} \label{subsec:UV_IR_SFR}

We produce a map of the SFR surface density in NGC 1365 following the procedure in \citet{Calzetti_2024}. This procedure describes the calibration of optical (H$\alpha$) tracers of unobscured star formation with IR data ($21\mu$m emission), serving as a tracer of dust-obscured star formation, to more robustly calculate the star formation in a given region \citep{Calzetti_2007}. 

We obtain a 24 $\mu$m map by calibrating the JWST MIRI F2100W (21$\mu$m) observations following the metholodology in \citet{Calzetti_2024}. 
In short, we select apertures within NGC 1365 with radii of 5$\arcsec$, and perform local background subtraction using an annulus background region surrounding the source aperture, with inner and outer radii of 5$\arcsec$ and 6.5$\arcsec$, respectively. 
Using the same regions, we similarly extract fluxes from a Spitzer 24$\mu$m image. We derive the calibration relation displayed in Equation \ref{eq:mips_jwst_calibration} below. 

\begin{equation}
    log(24\mu\textrm{m}) = log(21\mu\textrm{m}) - 0.73
    \label{eq:mips_jwst_calibration}
\end{equation}

We follow \citet{Calzetti_2024} equation 5 (reproduced below as Equation \ref{eq:calzetti2024_5}) to obtain the SFR surface density, adopting a scaling factor of $5.3\times(10^{-42})$, and a value for $b$ of $0.095 \pm 0.007$ \citep[see Equations 6 and 7; ][]{Calzetti_2007}. 

\begin{equation}
    \Sigma(SFR) \propto \Sigma(H\alpha_{corr}) = \Sigma(H\alpha) + b\Sigma(24)
    \label{eq:calzetti2024_5}
\end{equation}

\begin{figure}[h!]
\centering
\includegraphics[width=\columnwidth]{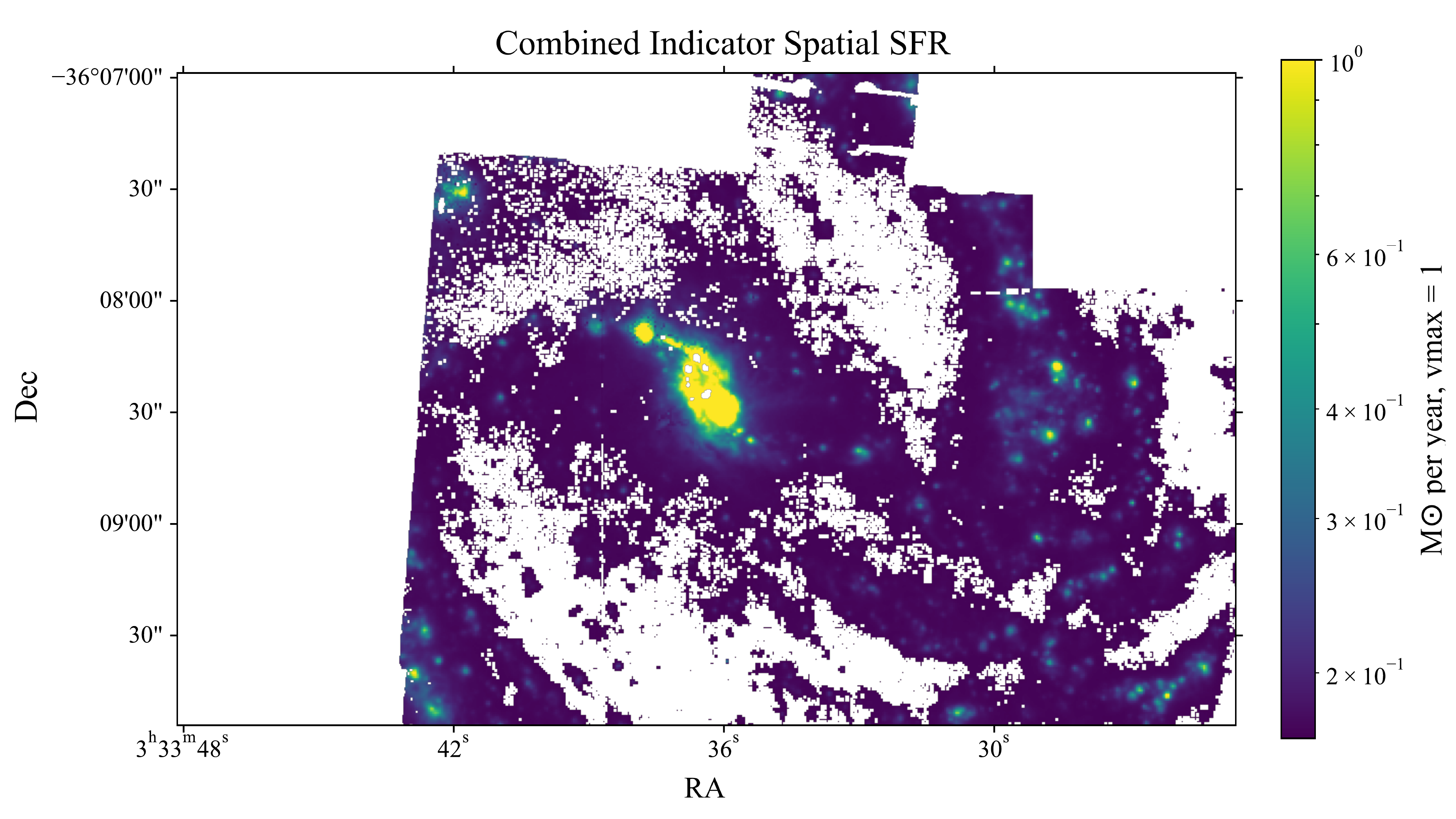}
\caption[width=\columnwidth]{SFR surface density map for NGC 1365 as calculated from JWST and optical H$\alpha$ data in Section \ref{subsec:UV_IR_SFR}. 
We present the SFR surface density in ~M$_\odot$~yr$^{-1}$~kpc$^{-1}$, and limit the plotted value at 1 ~M$_\odot$~yr$^{-1}$~kpc$^{-1}$ to show the structure in the spiral arms. As can be seen, the majority of star formation is concentrated within the nucleus, though there are regions of scattered heightened star formation in the lower left spiral arms.}

\label{fig:sfr_map}
\end{figure}

The resulting spatial SFR surface density map is presented in Figure \ref{fig:sfr_map}).

We note that we derive star formation rate surface densities ranging from $0.16 \leq$~M$_\odot$~yr$^{-1}$~kpc$^{-1}$$\leq 7.57$; however to better demonstrate structure outside of the central starburst, we plot a maximum value of M$_\odot$~yr$^{-1}$~kpc$^{-1}$ = 1 in Figure \ref{fig:sfr_map}. These star formation rate surface densities are in agreement with studies of larger samples of normal star forming galaxies  \citep{Teklu2020}.

\subsection{Spatially mapping dominant ionization source} \label{subsec:BPT_diagrams}
To explore the spatial extent of the AGN's impact on the host galaxy, we utilize multiple methods to identify the dominant source of emission per spaxel of the mosaic. One such diagnostic is the Baldwin-Philips-Terlevich (BPT) diagram \citep{BPT, kewley2001, kauffmann2003}. 

These diagnostics use optical emission line ratios ([\ion{N}{2}]/H$\alpha$, [\ion{O}{3}]/H$\beta$, [\ion{S}{2}]$\lambda6717,6713$/H$\alpha$, [\ion{O}{1}])/H$\alpha$) to identify the primary source of emission from a given region of ionized gas, based on the derived hardness of the ionizing radiation. Through the use of these diagnostics, a galaxy or region can be classified as being primarily SF-, AGN-, or LINER- (low ionization nuclear emission line region) dominated, or a composite of multiple sources of ionization \citep{kewley2001, kauffmann2003}.

\begin{figure*}[h!]
\centering
\includegraphics[width=0.90\textwidth]{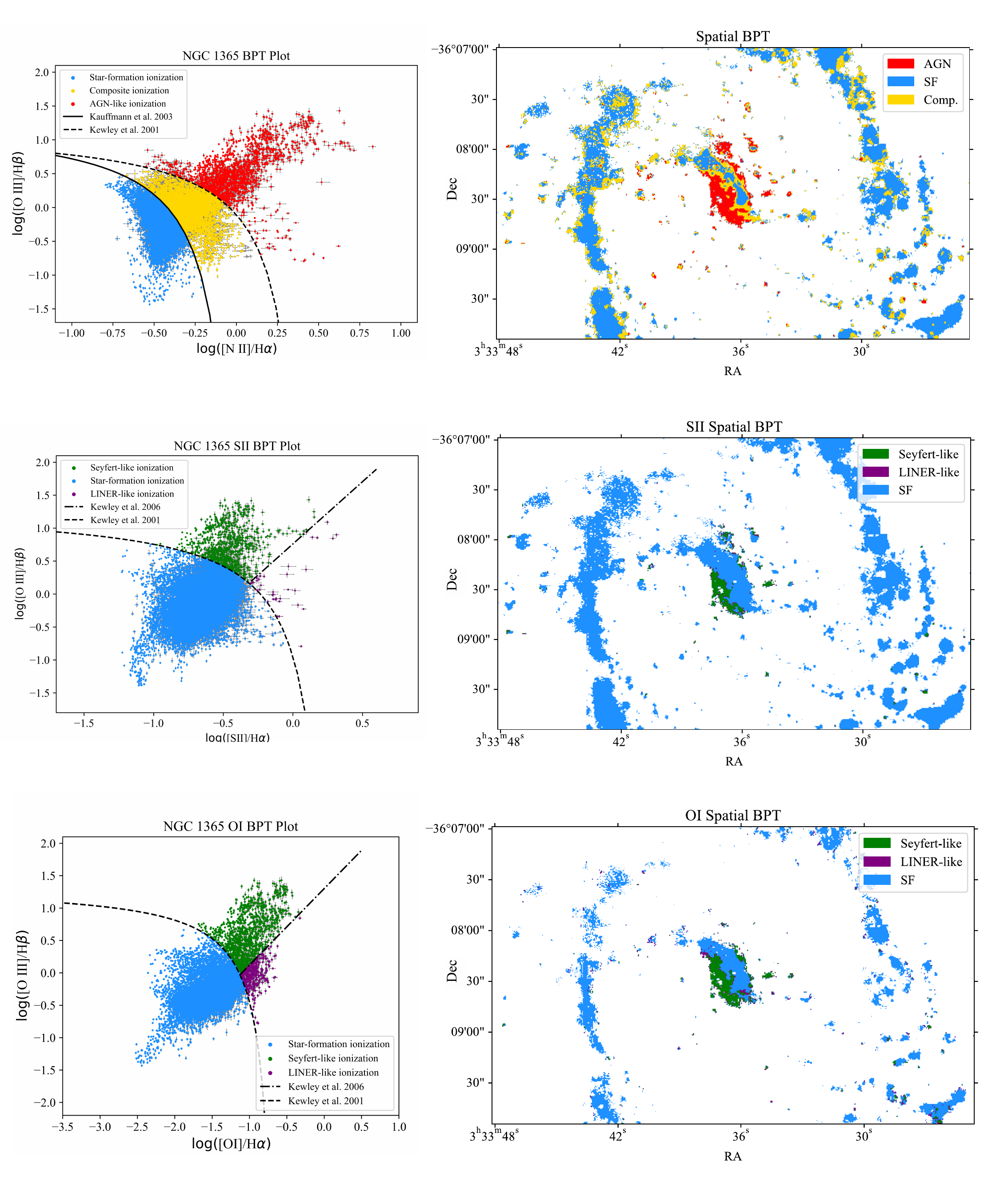}
\caption{The BPT diagrams (left) and maps (right) for each of the [\ion{N}{2}]/H$\alpha$ (top), [\ion{S}{2}]/H$\alpha$ (middle), and [\ion{O}{1}]/H$\alpha$ (bottom) diagnostics. Each spaxel in the map is represented by one scatter point in the BPT diagram. In the [\ion{N}{2}]/H$\alpha$ BPT diagram, we show the theoretical model for the upper-limit of star formation in \citet{kewley2001} (dashed black curve) and the empirical relation put forth in \citet{kauffmann2003} (solid black curve). The data have been masked per the method described in Section \ref{sec:data}. In all plots, blue represents emission suggestive of star formation. In the top plots, AGN-dominated emission is plotted in red, while spaxels displaying composite AGN/SF emission are denoted in yellow. In the bottom panels, Seyfert-like emission is plotted in green, while LINER-like emission is plotted in purple. As can be seen, the AGN-like emission is typically located in the nucleus but also remarkably appears in small clusters in the spiral arms. SF-dominant emission is predominantly seen in the spiral arms and in the starbust ring/dust lane in the nucleus. Composite regions appear on the borders of the dust lane in the nuclear region, as well as in significant patches along the star forming arms. Through these diagnostics, it appears that the AGN's region of impact is predominantly centered around the nucleus at a range of $\approx$3.6 kpc.
\label{fig:bpt_all}}
\end{figure*}

In Figure \ref{fig:bpt_all}, we utilize three BPT diagnostics to map the dominant source of ionization per spaxel in NGC 1365. 
Here, it provides a clear view of the strong nuclear starburst and dust lane across the center, and distinguishes this emission from surrounding AGN-dominated emission.  We note the ionization profile of the central region of the resultant spatial BPT map is well aligned with previous studies of the central region shown in \citet{Venturi2018MIRACLE}. Through these BPT diagnostics, the region of AGN-dominated emission, serving as a tracer of AGN-impacted regions, appears to be concentrated to the nuclear region, with a spatial extent of $\sim$3.6 kpc. 

We further note that in the BPT diagrams, a population of LINER-like points is seen partially in the [\ion{S}{2}] BPT and most clearly in the [\ion{O}{1}] BPT diagram, where it lies along the border of the SF-dominated and Seyfert-dominated emission surrounding the central AGN. The origin of this LINER-like emission is unclear with the current data, however it could align with additional sources finding LINER emission in the nuclear region of Seyfert 2 galaxies  \citep{Jingzhe2021}. Additionally, it could be attributed to the illuminated surface of an expanding pressure bubble \citep{stern2016} as it appears to align with a biconical outflow as traced by [\ion{O}{3}] (see discussion of outflows in Section \ref{subsec:kinematic_markers} for more details and shock activity in Section \ref{subsubsec:role_of_shocks}). To better understand the impact of potential shock activity in this region, additional data is needed in the near-infrared to access the H-band [\ion{Fe}{2}] shock diagnostics.

Finally, we note the presence of both significant composite regions and small clusters of AGN-like emission in the star-forming arms. To better understand how galaxy geometry may affect these observations, we note that the inclination of the galaxy is neither perfectly face- nor edge-on, with an inclination of 40-55$\degree$, where the northwest side is nearest \citep{Jorsater1995}. We discuss these regions in further detail, and potential explanations for this unique emission, in Section \ref{subsec:ionization source}.

As an alternative metric to identify the source of ionization with greater nuance, based on the spatially resolved emission throughout the galaxy, we derive the  ``AGN fraction" per spaxel of NGC 1365. The ``AGN fraction" is defined as the fraction of emission of a given spaxel that can be likely attributed to AGN activity. This diagnostic more finely indicates the spatial variation of ionizing radiation hardness, and offers another method by which to represent the magnitude of AGN effects on a given region.

\begin{figure*}
\includegraphics[width = 0.95\textwidth]{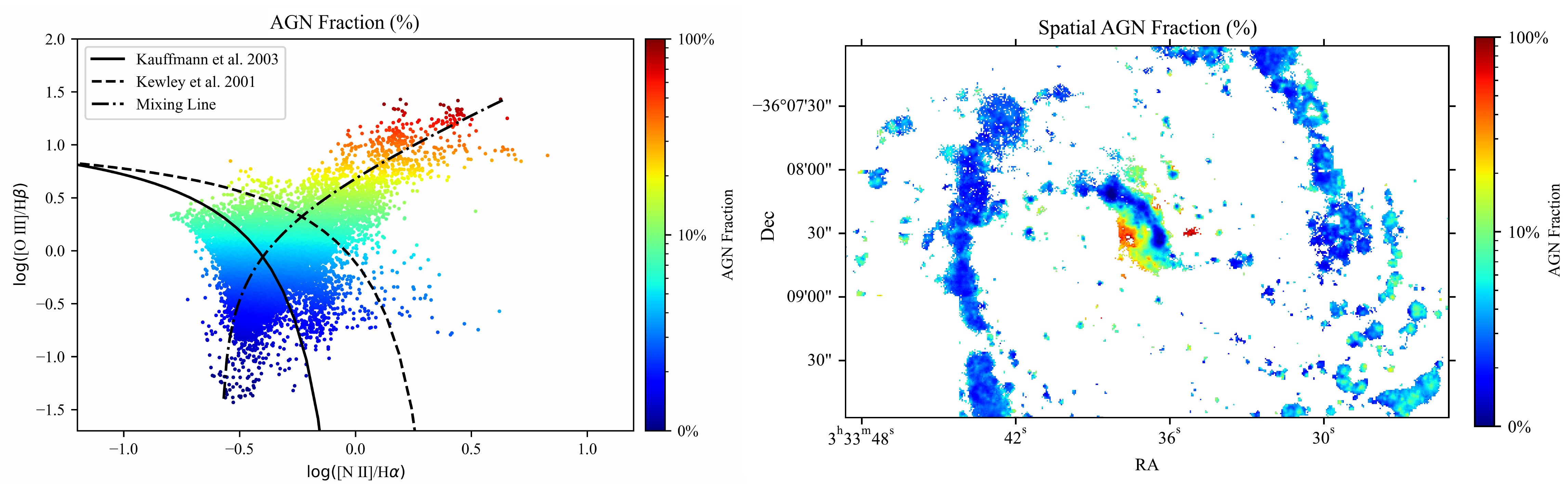}
\centering
\caption{Left: BPT diagram with the color bar denoting AGN fraction as defined in Section \ref{subsec:BPT_diagrams}. Right: Spatial map of AGN fraction. 
As can be seen, the AGN fraction similarly selects the unusual regions of AGN-like and composite emission in the spiral arms ($f_{\text{AGN}}$ $\approx$7-20$\%$), as well as an off-nuclear cluster of very high AGN-fraction spaxels ($f_{AGN}\sim{80\%}$). While the BPT diagnostics shown in Figure \ref{fig:bpt_all} display the regions of high ionization in the southeast spiral arms as scattered, this diagnostic shows a broader enhancement of potentially AGN-like emission, suggesting that the source of this emission could be more widespread than it would otherwise seem.
}
\label{fig:AGNfrac_w80_sfr}
\end{figure*}

We estimate the AGN fraction, $f_{\text{AGN}}$, for each spaxel following the procedure in \citet{Shin2019}, by first creating a linear space version of the scatter plot [\ion{N}{2}]/H$\alpha$ BPT diagram. We then select two basis points to represent 100\% star formation and 100\% AGN based on their locations on the BPT diagram representing the most extreme star formation and AGN classifications, and draw a linear mixing ratio relation between the two to denote the percent AGN fraction. We define that, for each spaxel on the linear space scatter plot, the nearest point along the mixing ratio line represents the percent AGN fraction of that spaxel. This relation is then translated back to log-log space, and spatially mapped on the galaxy in Figure \ref{fig:AGNfrac_w80_sfr} . 

The AGN fraction provides increased detail in comparison to the BPT, especially for regions high in AGN-fraction. In the nuclear region, areas of particularly high AGN fraction can now be identified, including an off-nuclear grouping that also appears in the central BPT in \citet{Gao2021}. As seen in Figure \ref{fig:AGNfrac_w80_sfr}, this source is shown to exhibit heightened [\ion{O}{3}] FWHMs ($\approx200-300$~km~s$^{-1}$), and lies within the outflow cone, suggesting that this emission can likely be attributed to AGN feedback. 

Additionally, we note a region in the southwest portion of the image, where a broad region of slightly heightened $f_{\text{AGN}}$  ($\approx$7-20$\%$) can be seen coincident with the region of scattered AGN-like and composite emission shown in the BPT maps (Figure \ref{fig:bpt_all}). This region displays $f_{\text{AGN}}$ values higher than the other star-forming areas ($f_{\text{AGN}}$  ($\approx$1-7$\%$)), and suggests that the source of the AGN-like and composite emission seen in the spatially resolved BPT maps may be more widespread than would appear through the BPT diagnostics alone. 
This unique emission in the spiral arms is further discussed in Section \ref{subsec:ionization source}.

\subsection{Kinematic Markers}
\label{subsec:kinematic_markers}
\begin{figure}[h!]
\includegraphics[width = 0.49\textwidth]{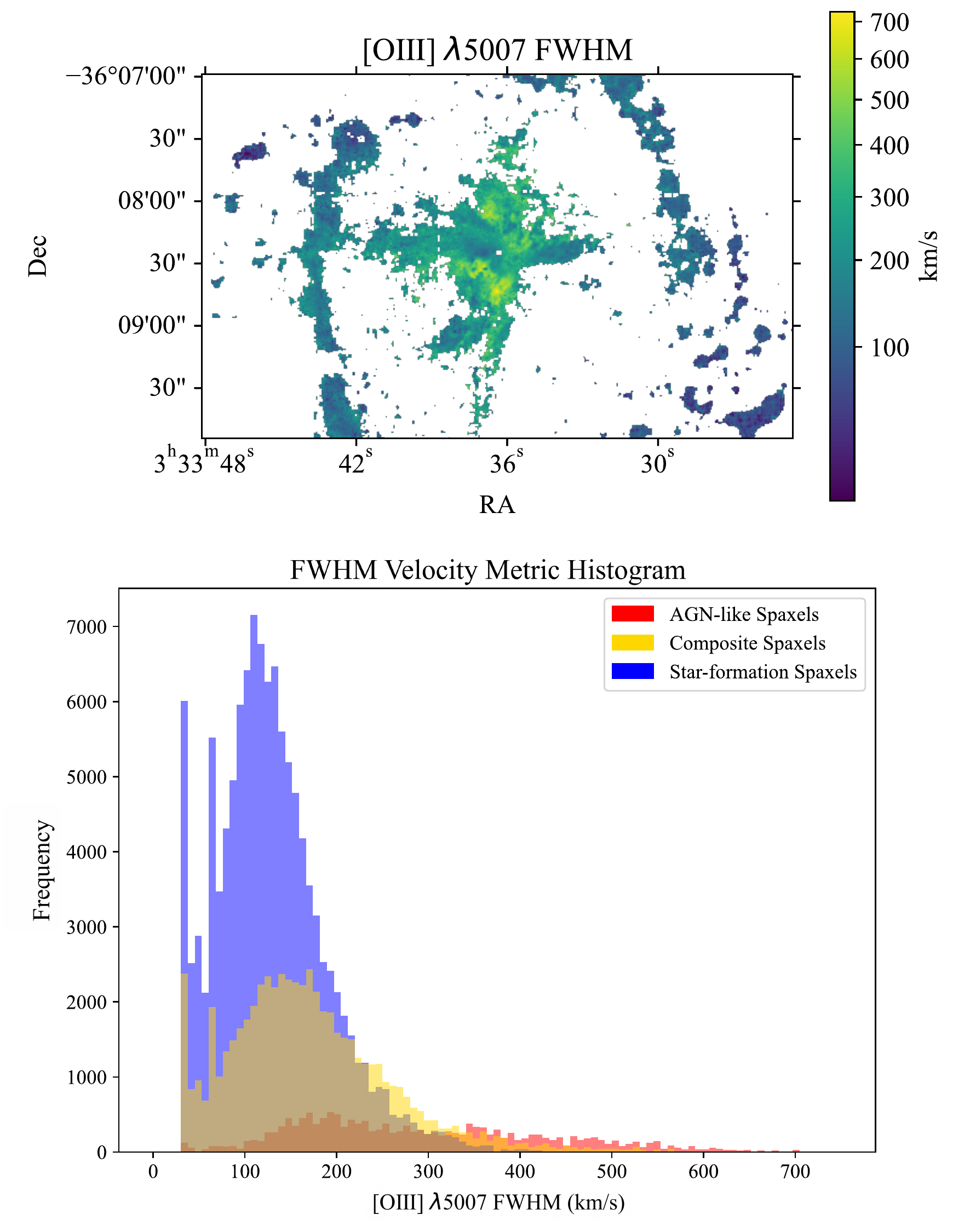}
\caption{Top: Map of [\ion{O}{3}]$\lambda5007$ FWHM. Here we see the peak velocities concentrated just to the northwest and southeast of the nuclear region, in line with a biconical outflow. 
Bottom: Histogram comparing the distribution of [\ion{O}{3}]$\lambda5007$ FWHM across BPT classifications \citep{kewley2001, kauffmann2003}, with AGN-dominated, composite, and SF-dominated spaxels shown in red, yellow, and blue, respectively. As can be seen, the three populations show distinct velocity distributions, with K-S statistical measures of 0.64 between the AGN and SF distributions, 0.39 between the Composite and AGN distributions, and 0.23 between the Composite and SF distributions. These results, each with p-values $< 0.0001$, demonstrate that each population is pulled from a different distribution. It can be seen that the distribution of velocity dispersion in the AGN-identified spaxels is significantly more broad, and the difference in population is confirmed with K-S testing. 
}
\label{fig:w80_sfr_hist}
\end{figure}

To assess the kinematic structure, we trace the distribution of [\ion{O}{3}]$\lambda5007$ bulk velocity FWHM across the galaxy in Figure \ref{fig:w80_sfr_hist}. 
Here, we note a clear view of the walls of the known outflow towards the northeast and southwest of the image. We can further compare with the [\ion{N}{2}]/H$\alpha$ BPT diagram in Fig. \ref{fig:bpt_all}. In doing so, we note that the spatial alignment of the composite regions in the star forming arm to regions of increased FWHM is mixed. For some regions, particularly on the outskirts of the star forming arms, increased FWHM and increased ionization align. However, in other regions, especially those internal to the star forming arm, the FWHM is not increased in composite regions. 

Additionally, we present a histogram of the FWHM of [\ion{O}{3}]$\lambda5007$ (Figure \ref{fig:w80_sfr_hist}), separated into three groups defined by their BPT classifications as calculated in section \ref{subsec:BPT_diagrams}. We note that the FWHM distribution of the full galaxy ranges from 31 to 728 km~s$^{-1}$, with averaged $68$\% confidence intervals of -21 km~s$^{-1}$ and +23 km~s$^{-1}$. The star formation categorized spaxels lie primarily within 30-360 km~s$^{-1}$, in alignment with typical FWHM values for the spiral arms of other massive galaxies \citep{Whittle1985}. The AGN-like categorized spaxels have a broader range, primarily from 50-720 km~s$^{-1}$, which reaches the ranges of AGN-driven outflows \citep{Harrison2014}, but also includes spaxels displaying lower velocities. As we only perform narrow line fitting in order to focus on the new addition of the star-forming arms, we expect not to capture all outflow components, and refer to the nuclear kinematic analysis in \citep{Venturi2018MIRACLE} for more detail. The composite categorized spaxels peak between these distributions, and we note that they spatially span areas of both the nuclear region and the star forming arms, including the higher-ionization regions of the star forming arms as seen in the BPT and AGN fraction diagrams (Figs. \ref{fig:bpt_all} and \ref{fig:AGNfrac_w80_sfr}). 
The three populations show distinct distributions. To test the significance of this difference, we performed a K-S statistical test, finding that each population in the histogram represents a distinct distribution, each with p $< 0.0001$, with K-S stat measures of 0.64 between the AGN and SF distributions, 0.39 between the Composite and AGN distributions, and 0.23 between the Composite and SF distributions. In particular, the velocity dispersion distribution in the star-formation ionization spaxels is significantly more narrow than in the BPT-identified AGN-like ionization distribution, and this is confirmed via K-S testing.

\section{Discussion} \label{sec:discussion}

\subsection{Analysis of BPT-identified ``AGN-like" high-ionization sources in the spiral arms} 
\label{subsec:ionization source}

\begin{figure*}
\includegraphics[width = 0.95\textwidth]{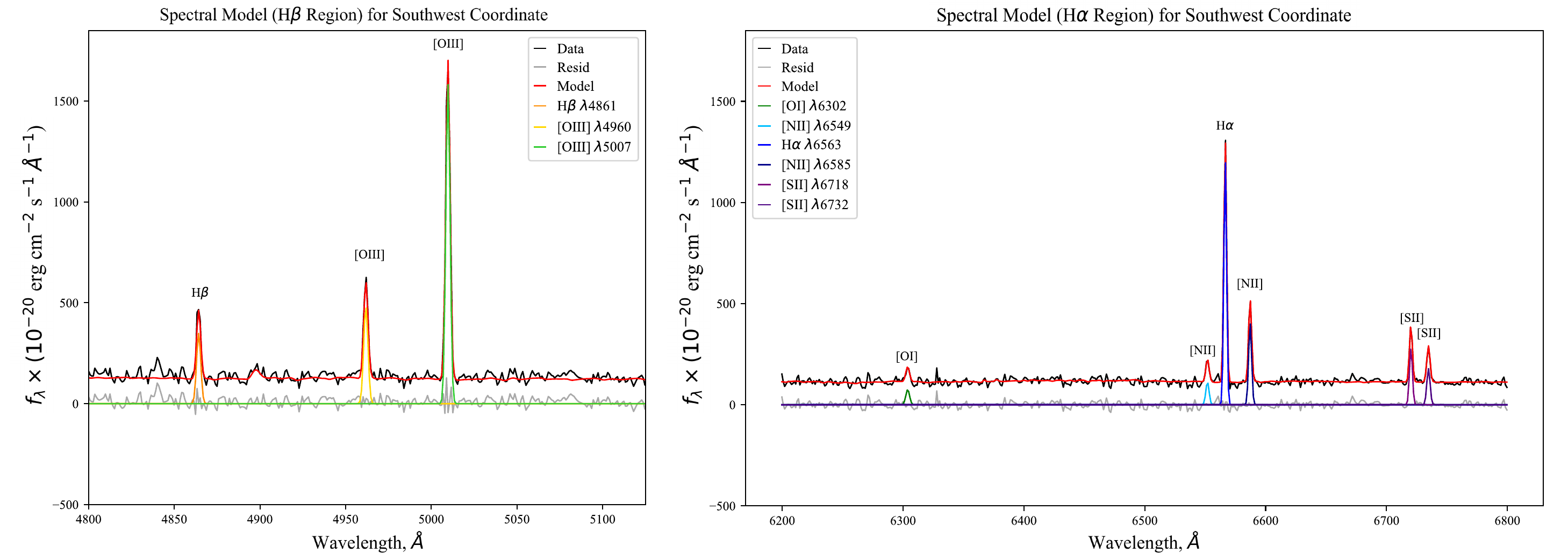}
\centering
\caption{Here, we show an example fit for a spaxel in the Seyfert/AGN-like region identified in the spiral arms in the southwest corner of the mosaic. The provided fit is created within the BADASS modeling process, and demonstrates the robust fit in this region. The fit is split into an H$\beta$-centered region in panel A and an H$\alpha$-centered region in panel B. 
\label{fig:exampleSeyfertcornerfit}}
\end{figure*}

Here we explore several potential origins of the  ``AGN-like" and composite emission in the star forming arms identified in Section \ref{subsec:BPT_diagrams}. 
Example spectra from this region can be found in Figure \ref{fig:exampleSeyfertcornerfit}, showing robust fits and confirming that the source of this discrepancy is not due to poor fits, but has a physical origin. 

\begin{figure}[h!]
\includegraphics[width = 0.49\textwidth]{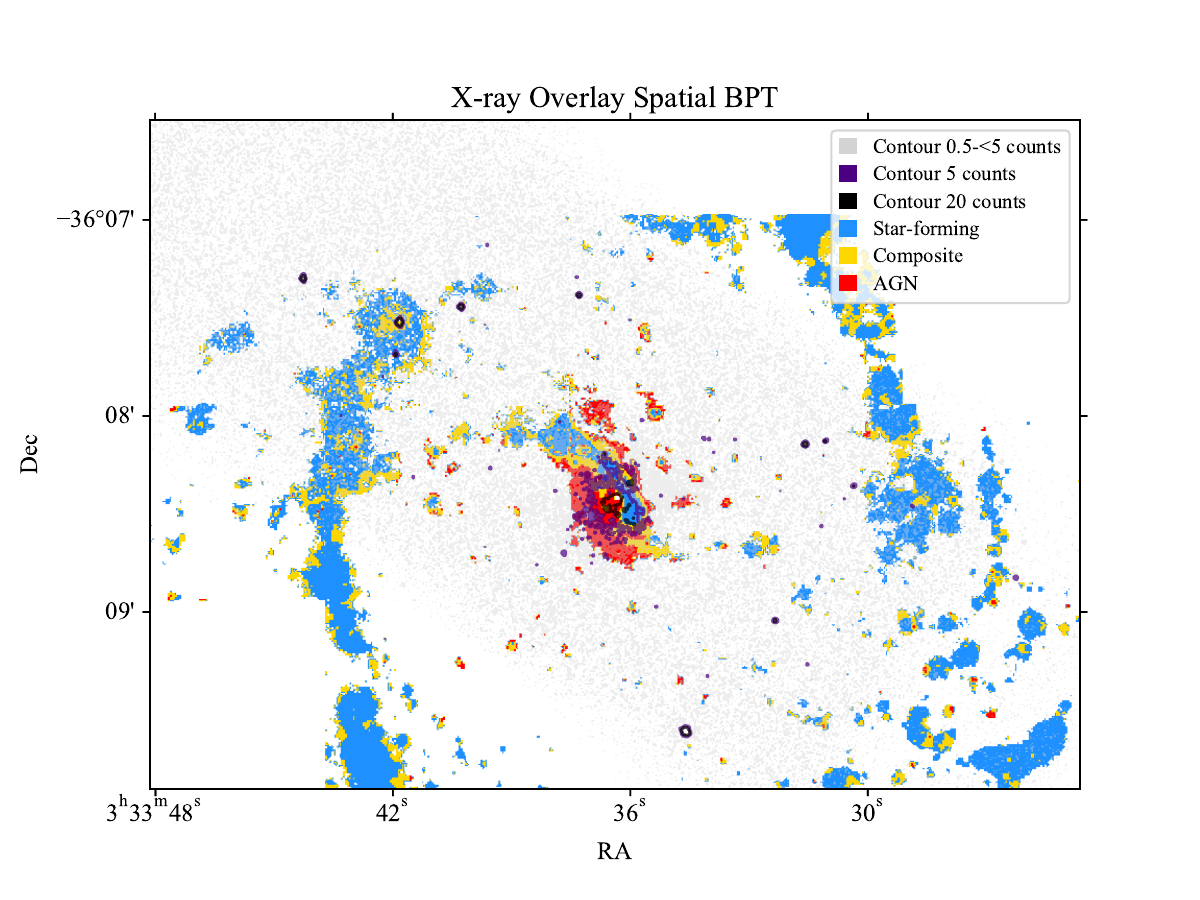}
\caption{Chandra contours (0.5-8 keV) overplotted on the [\ion{N}{2}]/H$\alpha$ BPT classification map from Figure \ref{fig:bpt_all}.
The gray contour level represents counts between 0.5-5, and is included to demonstrate the spatial coverage of the image in comparison with the optical MUSE coverage. The purple contour level corresponds to 5 counts, and the black contour level corresponds to 20 counts, to mark regions with significant X-ray emission. The strong X-ray emission is primarily confined to the nuclear region, with occasional peaks beyond the nucleus. As can be seen, there is no overlap between X-ray sources and the AGN-like regions in the spiral arms. 
\label{fig:bpt_xray}}
\end{figure}

We investigate the potential roles of other ionizing sources, including X-ray sources, supernovae, Wolf-Rayet stars and other extreme stellar populations, planetarey nebula, and shocks. We compare our classification and AGN-fraction maps to the survey of identified ultra-luminous X-Ray sources (ULXs) in NGC 1365 in \citet{Swain2023} and find no correlated ULXs. Further, a comparison with archival Chandra data of the galaxy shows no correlation between X-ray point sources and the BPT-identified sources (see Figure \ref{fig:bpt_xray}).

\begin{figure*}[t!]
\includegraphics[width = 0.95\textwidth]{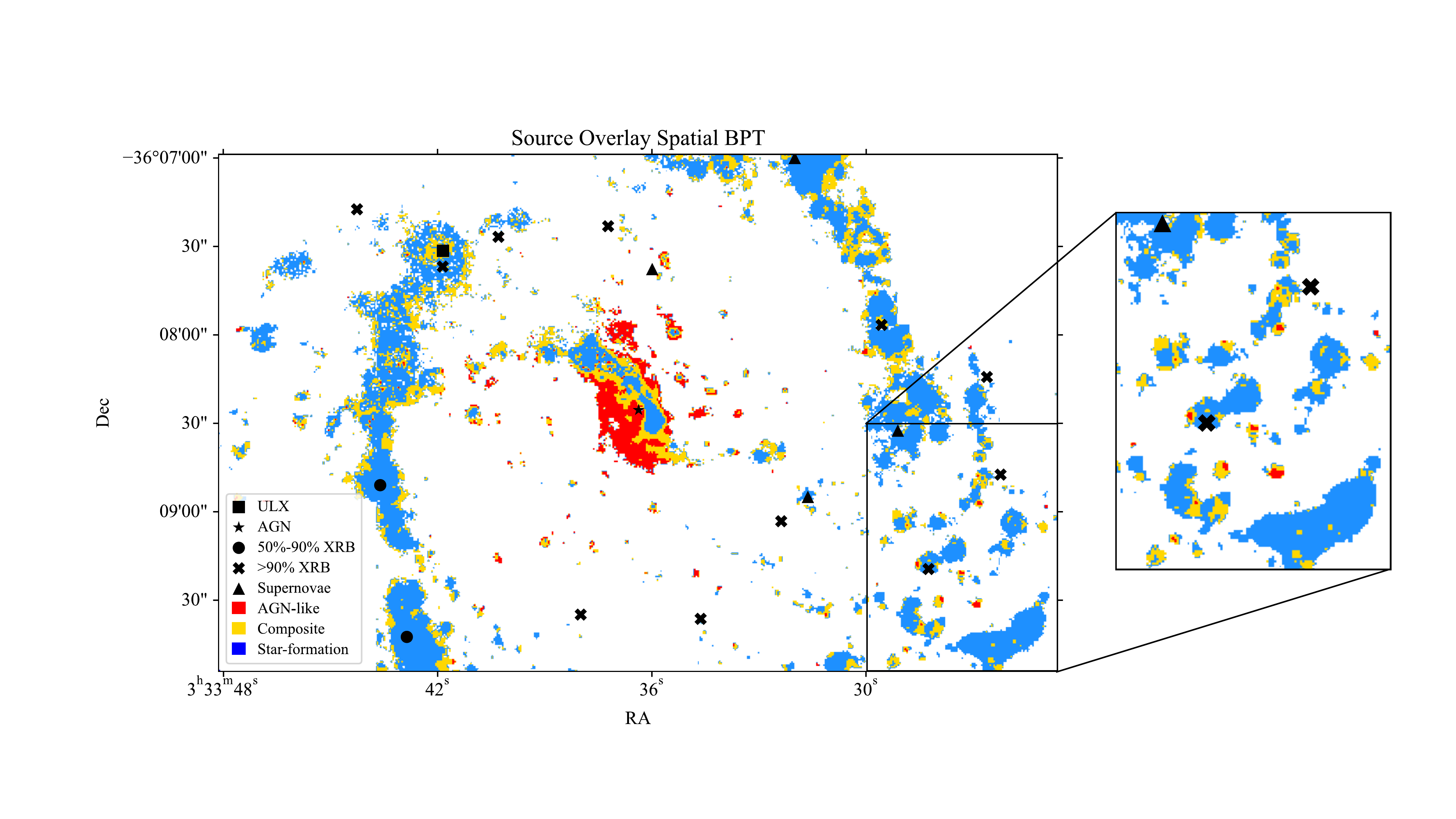}
\caption{Spatial [\ion{N}{2}]/H$\alpha$ BPT map from Fig. \ref{fig:bpt_all} overplotted with the locations of other astronomical sources, including objects identified with more than 90\% likelihood to be XRBs with X markers, objects identified with between 50\%-90\% likelihood to be XRBs with square markers, objects identified to be stars with star markers, and supernovae with triangle markers. As can be seen there are no overlapping sources with the AGN-like regions in the southwest corner of interest.
\label{fig:source_overlay}}
\end{figure*}

We further compare this data to other databases of previously identified sources in NGC 1365, including X-Ray Binaries (XRBs) from the XMM-Newton Source Analyzer \footnote{https://xmm-ssc.irap.omp.eu/claxson/xray$\_$analyzer.php \label{XMM Analyzer}} and supernovae from the Simbad database \footnote{https://simbad.u-strasbg.fr \label{SIMBAD}} in Figure \ref{fig:source_overlay}. We zoom in on the area in the southwest quadrant of the image where increased ionization is noted, to provide a closer look into the region. As can be seen, while there are sources in this quadrant, they do not directly correlate with the emission we are seeing.

On the example spectrum in Figure \ref{fig:exampleSeyfertcornerfit}, and other spectra from surrounding spaxels we find no broad lines or P Cygni profiles suggestive of supernova activity, nor the signature ``red" and ``blue bumps" indicative of Wolf-Rayet stellar populations. We additionally note that Wolf-Rayet stars are predominantly found in galaxies of high star formation ($>1$~M$_\odot$~yr$^{-1}$), which is not seen in this region, which displays star formation rate surface densities of $\sim 0.1$~M$_\odot$~yr$^{-1}$~kpc$^{-1}$ (see Figure \ref{fig:sfr_map}).

Planetary nebulae (PN) have also been shown to display AGN-like emission line ratios in the BPT diagram \citep{erzincan2025}, however the sources identified here reach [\ion{O}{3}] luminosities of $10^{37-38}$~erg~cm$^{-1}$, which is at or above the brightest end of the planetary nebula [\ion{O}{3}] luminosity function ($10^{36-37}$~erg~s$^{-1}$), suggesting that the emission is likely not attributable to PN.

As these sources were identified as regions of heightened AGN-like emission through the diagnostics in Section \ref{subsec:BPT_diagrams}, we also investigate the potential explanation for this emission as coming from an AGN-like source, whether via a unique view into the central engine, AGN-driven feedback, or the identification of an accreting off-nuclear massive black hole.

The sources in this region exhibit [\ion{O}{3}] luminosities of $10^{36-37}$~erg~s$^{-1}$. If we assume the bolometric correction factors for $L_{\textrm{[O~III]}}$ from \citet{lamastra2009} continue to the luminosities seen here, we find bolometric luminosities of $\sim10^{38-39}$~erg~s$^{-1}$. This would imply a 100-1,000~M$_\odot$ black hole accreting far below its Eddington limit, rendering an off-nuclear AGN explanation for the source of emission in these sources unlikely.

We instead explore the potential for AGN feedback as an explanation for this emission. As can be seen in Figure \ref{fig:w80_sfr_hist}, the average [\ion{O}{3}] FWHM in this region is extremely narrow ($<100$~km~s$^{-1}$), with no additional outflow component seen in visual inspection, and clearly separated from the outflow cones seen from the southeast to the northwest.

We additionally present a map of the Balmer decrement (H$\alpha$ / H$\beta$), used as a tracer to demonstrate regions with significant reddening from dust, and thus potentially to trace clumping of gas and dust due to AGN-driven outflows. Here, we note heightened Balmer decrements along the cone of the outflow (to the northwest and southeast), and lower values outside of this cone. This may further suggest that this region of AGN-like ionization in the southwest corner is not driven by the AGN.

\begin{figure}[h!]
\includegraphics[width = 0.49\textwidth]{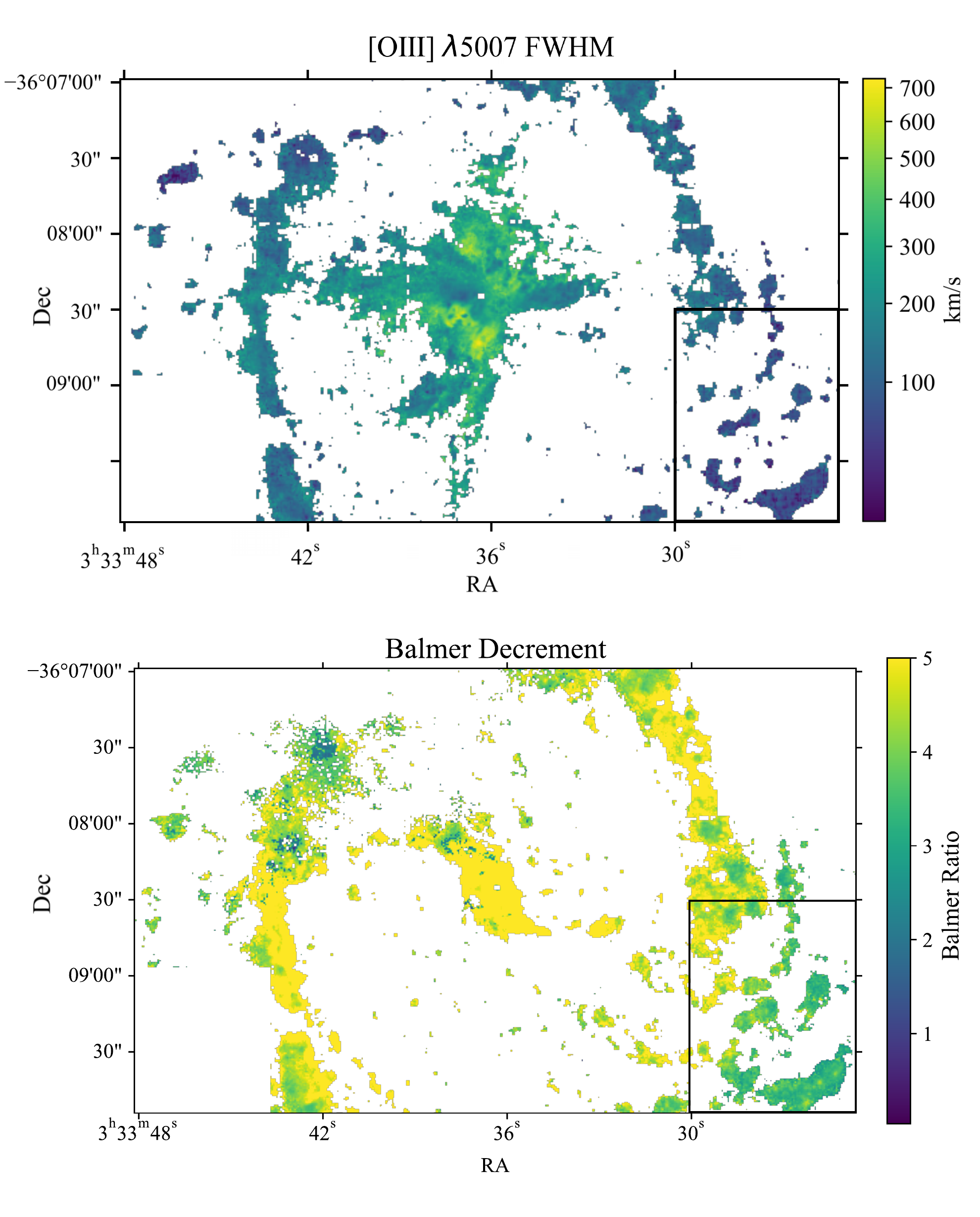}
\caption{Top: We again show the [\ion{O}{3}] $\lambda$5007 FWHM, here noting the region of high ionization in the Southwest star forming arms. Bottom: Here, we show the Balmer decrement map (ie. H$\alpha$ / H$\beta$). For clarity, we present it with a maximum value of 5 (the maximum value in the star forming arms), to prevent overshadowing by high Balmer decrements along the central dust lane, noting that in the nuclear region values can reach up to approximately 25. We note significant reddening along the central outflow cone, and lower values  outside of the outflow cone. 
\label{fig:balmerdec}}
\end{figure}

As the source of this emission appears to be unrelated to the central AGN, we now investigate the potential for the role of shock heating in creating this emission.

\subsection{The Role of Shock Heating} 
\label{subsubsec:role_of_shocks}
\begin{figure*}[h!]
\includegraphics[width = 0.85\textwidth]{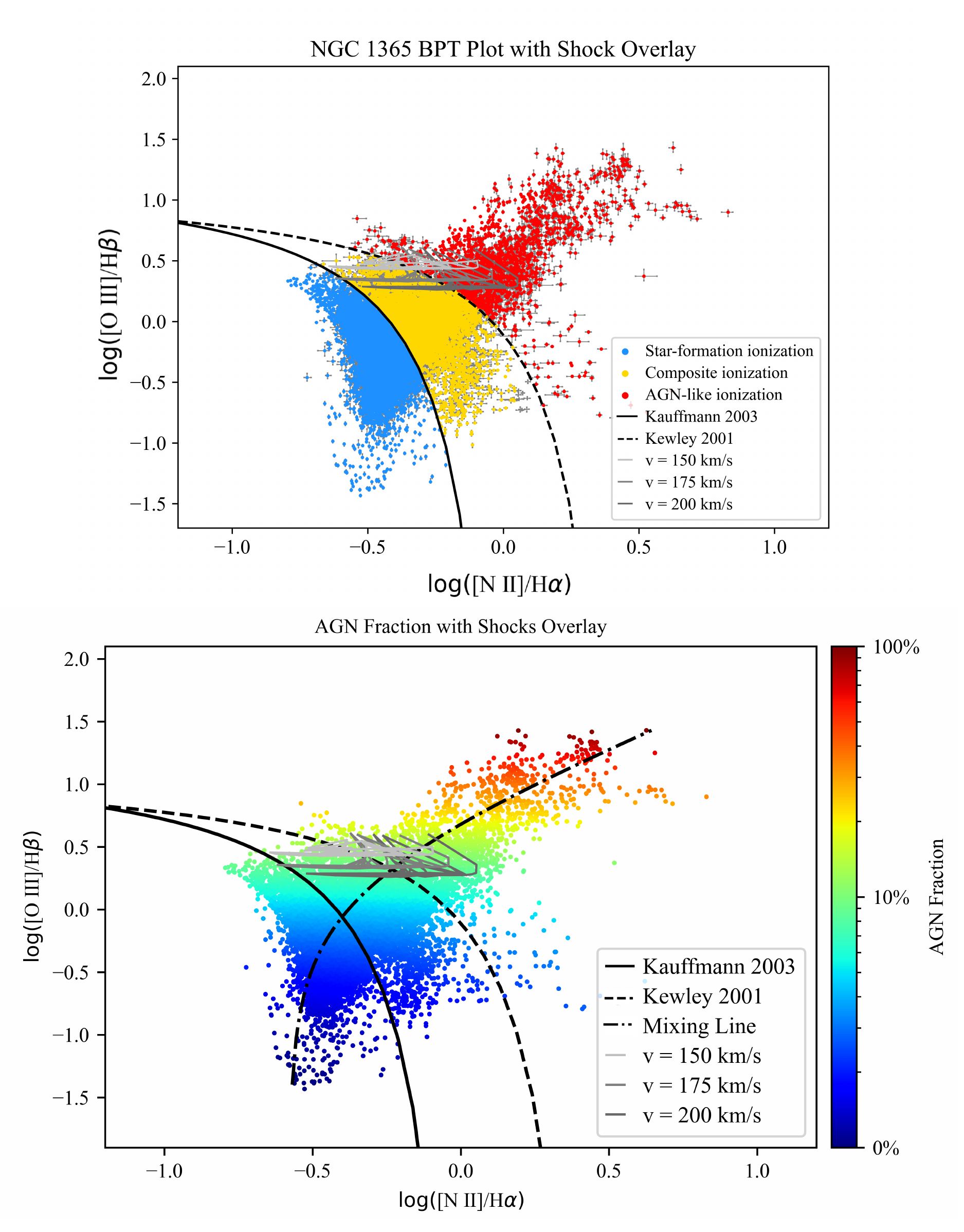}
\centering
\caption{Top: Shock models, gridded over density ($n_H =  0.1-10$~cm$^{-3}$) and magnetic field strength ($B=0.001-10~\mu$G), overplotted on the [\ion{N}{2}]/H$\alpha$ BPT diagram. Shock velocities ranging from $150-200$~km~s$^{-1}$ are denoted in grey, with higher velocities plotted in darker shades of grey. These models assume Solar metallicity.  Bottom: Same as above, but overplotted on the AGN Fraction diagram from Fig. \ref{fig:AGNfrac_w80_sfr}. As can be seen, the parameter space covered by this shock model significantly overlaps with the AGN Fractions produced by the spaxels in the southwest corner of BPT, where clumps of AGN-like emission are located. Additionally, we note that shocks alone can reproduce the observed emission, without the need for an AGN component.
\label{fig:shocks_NIIoverlay}}
\end{figure*}

Finally, we explore the potential for this emission to be driven by shocks. 
In Figure \ref{fig:shocks_NIIoverlay}, we overplot shock models from \citet{Morisset2015_3MdB}, referencing \citet{Allen2008}, using the online 3MdB repository \footnote{https://sites.google.com/site/mexicanmillionmodels/}. We select the Solar metallicity model based on the metallicities calculated in \citet{Ho_2017} and a velocity of 150-200 km~s$^{-1}$, based on the largest FWHM in this area calculated in \ref{fig:w80_sfr_hist}. We compute the resultant emission line ratios for the full grid of densities (0.1-10 cm$^{-3}$) and magnetic field strengths (0.001-10 $\mu$G) explored. We overplot this grid on the [\ion{N}{2}]/H$\alpha$ BPT diagram and the AGN Fraction diagram, to investigate if the emission from the model shares the parameter space of the unexpectedly high-ionization regions. 

As seen in Figure \ref{fig:shocks_NIIoverlay}, the shock model parameter space overlaps significantly with the region of the AGN Fraction diagram produced by the spaxels in the southwest corner of the [\ion{N}{2}] BPT diagram, including regions identified as AGN-like ionization. The parameter space additionally overlaps with regions of the composite BPT spaxels, indicating that shocks remain a possible explanation for the source of this emission, and we confirm that velocities needed to produce these shocks (150 - 200 km~s$^{-1}$ are found in this region of the galaxy, and are higher, but reasonable, values for velocity in the star-forming arms of massive galaxies \citep{Whittle1985}. 

Additionally, to better understand the potential source of shocks in this region, or other possible causes for this increased ionization, we note the necessity for spectral coverage of this region in near- and mid- infrared, to access other indicators of shocks and obscured activity. Shock diagnostics accessible in the near-infrared could help in confirming the presence of shocks in this region of higher-than-expected ionization \citep[e.g.,][]{Bohn2024_GOALS, Storchi_2009, Riffel2013, Inami_2013}, while a stronger understanding of the obscuration in this region is needed to evaluate if dust could play a significant role in obscuring detection of an ionization source within this region.

\subsection{Connections to Previous Studies}
\label{subsec:implications}
This work provides a detailed, spatially-resolved look into a common source  of ambiguity when trying to determine the dominant source of ionization in a system.
Ambiguity in emission-line diagnostics are common problems for determining the ionization profiles of galaxies, and this uncertainty is especially prevalent when considering unresolved sources \citep[see][section 7, for a review]{Kewley2019}. This ambiguity has significant implications to the compilation of a robust census of AGN activity and a clearer understanding of the role of AGN activity in galaxy evolution. Accurately disentangling the ionization sources in a galaxy is important to better understand the spatial scales on which different sources of potential feedback effects are active, and thus this ambiguity presents challenges for understanding galaxy ionization structures. 

For example, when comparing the AGN fractions found in large samples, surveys having higher angular resolution have found a larger population of AGN, particularly lower-mass and obscured AGN, when compared to surveys using lower resolution data \citep[e.g.,][]{Bridge_2016}. This discrepancy suggests that the AGN signature was overshadowed by emission from the host galaxy when considering the integrated spectrum, and that alternate metrics to traditional emission line ratios, or combinations of alternate metrics, may increase our effectiveness in identifying lower-mass AGN. 

Additionally, a similar ambiguity between AGN and shock-driven emission in the composite region of the BPT diagram in NGC 1365 has been seen in surveys of (Ultra) Luminous Infrared Galaxies (ULIRGs/LIRGs), where shock activity has been shown to replicate BPT-identified ``composite" emission without the need for an AGN component \citep[e.g][]{Rich2014_ULIRG}. This result further underscores that traditional emission line diagnostic interpretation may not be applicable in all scenarios, and that care must be taken in the interpretation of unresolved spectra to not over- or under-estimate the contribution of AGN activity.

This ambiguity is further exacerbated in dwarf galaxies, where the central AGN tends to be less massive, and therefore less luminous and more easily overshadowed by the light of active star formation using traditional optical diagnostics \citep{trump2015}. In fact, spatially resolved studies using the SDSS/MaNGA survey have uncovered significantly higher AGN fractions than traditional surveys \citep[$\sim20$\% vs. $\sim1$\%;][]{Mezcua2020_dwarf1, Mezcua2024_dwarf2}. As the occupation fraction of AGN in local dwarf galaxies is a critical parameter for constraining black hole seed models \citep[see][for a review]{greene2020}, this discrepancy has significant implications to our understanding of the formation and early evolution of supermassive black holes at high redshift.

The emission structure seen here in NGC 1365, especially concerning the region of higher than expected ionization in the SW star forming arms, adds to the significant literature investigating the ambiguity in these diagnostics, identifying additional possibilities for the complexities of ionization sources not represented in low-resolution data. Without the spatial resolution of the MUSE data, this region could be obscured or misidentified as being driven by the central AGN, discounting a significant source of ionization in the galaxy and obfuscating the complex picture of ionization contributing to the galaxy's evolution. 

\subsection{Implications for Future Studies}

This work underscores the need for careful interpretation of BPT analyses of observations that probe astrophysical systems at low-spatial resolution. This is particularly important for better understanding how we characterize the sources of ionizing radiation in objects at high-$z$. Since its commissioning, \textit{JWST} has revolutionized our understanding of galaxy and SMBH growth in the early Universe. It has in part done this through the discovery of an anomalous yet ubiquitous population of ``little red dots'' (LRDs) \citep[e.g.,][]{matt24lrd,koce25lrdsample}. LRDs are $z>4$ sources that are defined by their point-like morphologies and V-shaped spectral energy distributions (SEDs). The physical nature of LRDs is not yet fully understood. Their point-like morphologies and often presence of broad lines in their spectra suggest they may be AGN \citep[e.g.,][]{green24uncover_broadline_lrd,hvid25_rubies-bl-lrds}, yet their apparent X-ray weakness is either indicative of super-Eddington accretion or absorption by broad line emitting or ionized gas \citep[e.g.,][]{lamb24superEdd_lrd,maio25chandra_blr_lrd,degr25black_hole_star}. It is also possible that that their SEDs may be dominated by stellar processes, as also supported by their X-ray weakness and observed frame mid-IR luminosities \citep[e.g.,][]{pere24smiles_miri_lrd,bagg24starburst_broadline_lrd}. 

Emission line diagnostics would provide powerful constraints on the nature of LRDs, but they have been challenging to interpret due to the fact that other star forming high-$z$ galaxies have been shown to have high-ionization parameters and low-metallicities \citep[e.g.,][]{trum23_highz-emline-diag}. This makes local calibration of emission line diagnostics unreliable for these sources. Additionally, the unresolved morphologies of LRDs make them especially prone to misidentification via emission line flux ratios since there is no spatial context for the origin of the emission. The work being presented in this paper showed hard ionized emission originating from heavily star forming regions away from the nucleus of NGC 1365. The compact nature of LRDs makes it impossible to distinguish between hard ionization arising from AGN or extreme star formation on the basis of location in the galaxy the same way that it was possible to do for the ionization sources in NGC 1356. This emphasizes that emission line diagnostics such as the high-$z$ calibrated ``OHNO'' diagram \citep[e.g.][]{trou11high-z-bpt}  should primarily be used to identify the presence of highly ionized emission, not to infer the nature of the ionizing source in compact, high-$z$ sources.

The identification of BPT-identified ``AGN-like" and ``composite" emission in the spiral arms of NGC 1365 has further implications for the study of local dwarf galaxies. A complete census of AGN activity in dwarf galaxies, and particular in low metallicity ``local analog" dwarf galaxies, is a critical step towards constraining black hole seed models, however the traditional diagnostics to identify AGN activity in massive galaxies tend to lose efficacy in this regime \citep[see][and references therein]{cann2018}. This has resulted in a growing sample of AGN candidates identified in dwarf galaxies with limited or no confirmation from multi-wavelength diagnostics \citep[discussed in e.g.,][]{Latimer_2021, Mingozzi2025_Futuredwarf}. The presence of regions in the spiral arms of NGC 1365 showing emission traditionally indicative of AGN activity further emphasizes the need for caution in interpreting such results, as there remains significant uncertainty in the potential for non-AGN processes to mimic traditional AGN signatures.

\section{Conclusion} \label{sec:conclusion}

We present a high-spatial resolution analysis of the ionization and kinematic properties of NGC 1365 at unprecedentedly large physical scales. Through the use of multi-wavelength data, we present maps of the ionization state, bulk velocity kinematics, star formation surface density, and make comparisons to prior analyses. We use two strategies to spatially separate the ionization sources in the star forming arms, the BPT diagram and AGN fraction, and find that the BPT diagram produces three populations that display distinct FWHM velocity distributions.
Moreover, we identify regions of high-ionization emission in the star forming arms of the galaxy. In particular, regions displaying composite and AGN-like emission line ratios as classified via the BPT diagram have been identified, and the source of their emission has been explored. Through an analysis of potential sources, we find that shock heating models aligned with the top FWHM velocity measures in the area and metallicity of the galaxy have the potential to produce the observed emission, however additional observations in the near-infrared are necessary to confirm this hypothesis.
This result highlights the necessity of spatially-resolved studies and the need for caution when working with unresolved data, as this high ionization could have been solely attributed to AGN activity in globally averaged measurements, failing to account for a significant driver of energy transfer in the spiral arms, with implications for the interpretation of lower resolution data. 

\begin{acknowledgments}

Based on observations made with ESO Telescopes at the La Silla Paranal Observatory under programme ID 094.B-0321; 1100.B-0651. 

The scientific results reported in this article are based in part on data obtained from the Chandra Data Archive. This research has made use of software provided by the Chandra X-ray Center (CXC) in the application package CIAO.

The material is based upon work supported by NASA under award number 80GSFC24M0006.

Resources supporting this work were provided by the NASA High-End Computing (HEC) Program through the NASA Center for Climate Simulation (NCCS) at Goddard Space Flight Center.

This research made use of Photutils, an Astropy package for
detection and photometry of astronomical sources \citep{photutils}

This material is based upon work supported by the National Science Foundation Graduate Research Fellowship Program. Any opinions, findings, and conclusions or recommendations expressed in this material are those of the author(s) and do not necessarily reflect the views of the National Science Foundation.

\end{acknowledgments}

\facilities{VLT (MUSE), JWST (MIRI), Chandra (ACIS), Spitzer (MIPS 24$\mu$)}

\software{Astropy \citep{2013A&A...558A..33A,2018AJ....156..123A,2022ApJ...935..167A}, Source Extractor \citep{1996A&AS..117..393B}, BADASS \citep{BADASS}, Matplotlib \citep{matplotlib}, Photutils \citep{photutils} }

\bibliography{sample701}{}
\bibliographystyle{aasjournalv7}

\end{document}